%% file: main.tex
\documentclass[letterpaper]{article} % DO NOT CHANGE THIS
\usepackage{aaai2026}  % DO NOT CHANGE THIS
\usepackage{times}  % DO NOT CHANGE THIS
\usepackage{helvet}  % DO NOT CHANGE THIS
\usepackage{courier}  % DO NOT CHANGE THIS
\usepackage[hyphens]{url}  % DO NOT CHANGE THIS
\usepackage{graphicx} % DO NOT CHANGE THIS
\usepackage{natbib}  % DO NOT CHANGE THIS AND DO NOT ADD ANY OPTIONS TO IT
\usepackage{caption} % DO NOT CHANGE THIS AND DO NOT ADD ANY OPTIONS TO IT
\usepackage{algorithm}
\usepackage{algorithmic}

\usepackage{newfloat}
\usepackage{listings}
\DeclareCaptionStyle{ruled}{labelfont=normalfont,labelsep=colon,strut=off} % DO NOT CHANGE THIS
\floatstyle{ruled}
\newfloat{listing}{tb}{lst}{}
\floatname{listing}{Listing}
\nocopyright  % -- Your paper will not be published if you use this command
\title{L3AC: Towards a Lightweight and Lossless Audio Codec}

\author{
    Linwei Zhai\textsuperscript{\rm 1},
    Han Ding\textsuperscript{\rm 1},
    Cui Zhao\textsuperscript{\rm 1},
    Fei Wang\textsuperscript{\rm 1},
    Ge Wang\textsuperscript{\rm 1},
    Zhi Wang\textsuperscript{\rm 1},
    Wei Xi\textsuperscript{\rm 1}
}
\affiliations{
    \textsuperscript{\rm 1}Xi'an Jiaotong University, China\\

}

\usepackage{bibentry}
\input{command}

\begin{document}

    \maketitle

    \begin{abstract}
        Neural audio codecs have recently gained traction for their ability to compress high-fidelity audio and provide discrete tokens for generative modeling.
        However, leading approaches often rely on resource-intensive models and complex multi-quantizer architectures, limiting their practicality in real-world applications.
        In this work, we introduce \system, a lightweight neural audio codec that addresses these challenges by leveraging a single quantizer and a highly efficient architecture.
        To enhance reconstruction fidelity while minimizing model complexity, \system explores streamlined convolutional networks and local Transformer modules, alongside \textit{TConv}—a novel structure
        designed to capture acoustic variations across multiple temporal scales.
        Despite its compact design, extensive experiments across diverse datasets demonstrate that \system matches or exceeds the reconstruction quality of leading codecs while reducing computational overhead by an order of magnitude.
        The single-quantizer design further enhances its adaptability for downstream tasks.
        The source code is publicly available at \url{https://github.com/zhai-lw/L3AC}.
    \end{abstract}

    \input{body/intro}
    \input{body/related_works}
    \input{body/model}
    \input{body/experiments}
    \input{body/conclusion}

    \input{body/others}

    \clearpage
    \bibliography{refs}

    \input{appendix/all}

\end{document}

%% file: command.tex
\usepackage{makecell}
\usepackage{xspace}

\usepackage{color, colortbl} % (if used in text)

\usepackage{microtype}
\usepackage{subfigure}
\usepackage{booktabs}
\usepackage{subcaption}

\usepackage{amsmath}
\usepackage{amssymb}
\usepackage{mathtools}
\usepackage{amsthm}

\usepackage[capitalize,noabbrev]{cleveref}

\newcommand{\appendixref}[1]{Appendix~#1}

\newcommand{\ie}{\emph{i.e.}\xspace}
\newcommand{\eg}{\emph{e.g.}\xspace}
\newcommand{\etc}{\emph{etc}\xspace}

\makeatletter
\newcommand{\system}{%
    L3AC%
    \@ifnextchar\_{}{%
        \@ifnextchar,{}{%
            \xspace}}%
}
\makeatother

%% file: body/intro.tex
\section{Introduction}

Recent advances in neural audio codec technologies have markedly improved the compression and reconstruction of high-fidelity audio.
Unlike traditional audio codecs, systems based on deep neural networks not only compress audio efficiently but also produce discrete codes that can be utilized as tokens in sound language modeling (LM) \cite{CodecSuperB2024}.
This dual functionality underscores their critical importance in modern audio processing tasks.
By integrating tokenized outputs of neural audio codecs with language models, a seamless bridge is formed between audio compression and generative language modeling, opening the door to a variety of novel applications.

Despite these benefits, many state-of-the-art (SOTA) neural codecs \cite{SoundStream2021, HiFi-Codec2023X, FunCodec2024, encodec2023, dac2024} rely on complex, multi-quantizer architectures to achieve their high performance in compression and reconstruction.
These systems typically employ multiple quantizers arranged hierarchically, each capturing different levels of detail in the audio signal.
While this structure enhances reconstruction fidelity, it introduces two significant limitations.
First, the resultant hierarchical token streams necessitate customized aggregation techniques to support downstream models \cite{copet2023,WavTokenizer2025}, complicating both training and inference.
Second, as the number of quantizers increases, inconsistencies can emerge in the generated tokens.
These inconsistencies make it challenging for language models to reliably predict subsequent tokens \cite{liu2024X}.

Although some multi-quantizer frameworks can function as a single quantizer by utilizing only one of their components, this often sacrifices reconstruction quality.
In response, recent efforts \cite{Single-codec2024, TAAE2025, WavTokenizer2025} have explored single-quantizer neural codecs.
These codecs offer a more streamlined alternative, yet they show only modest improvements in reconstruction quality over their multi-quantizer counterparts and often rely on resource-intensive architectures, which limits their practical applicability.

\begin{figure}[t]
    % \vskip 0.2in
    \begin{center}
        \centerline{\includegraphics[width=.85\columnwidth]{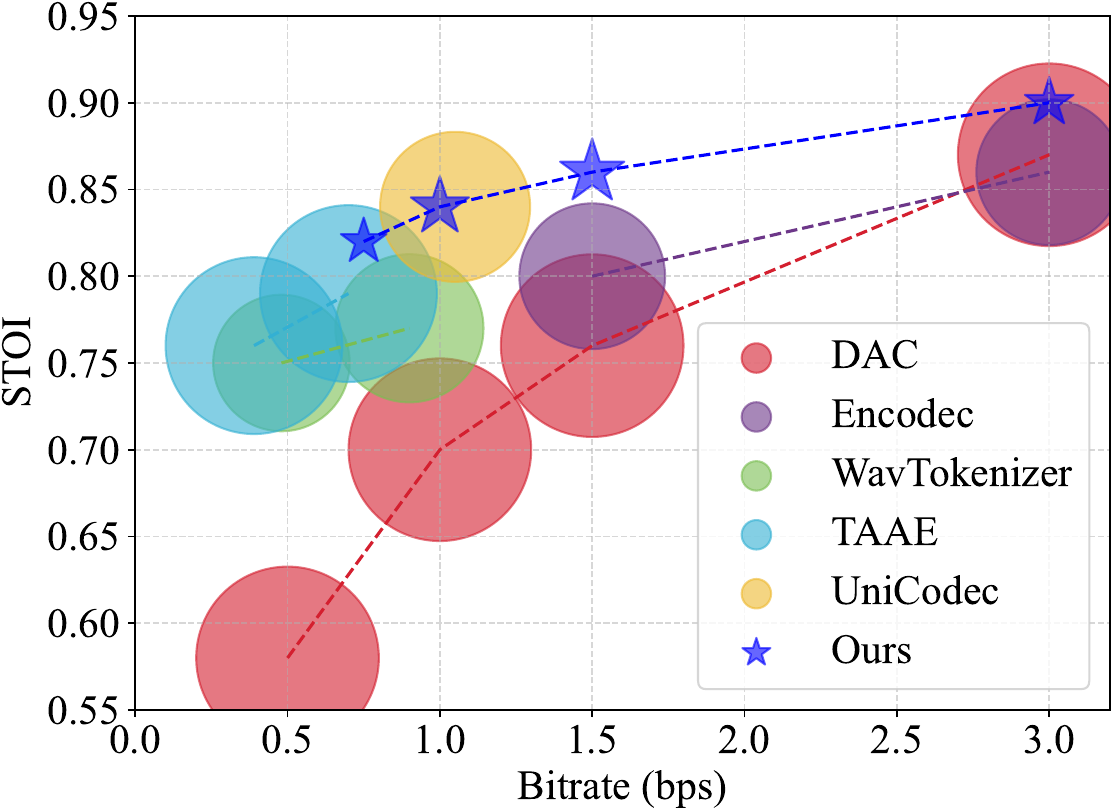}}
        \caption{
            Comparison of various audio codecs.
            The y-axis (STOI) indicates the quality of the reconstructed audio, where higher values denote better performance.
            The x-axis (Bitrate) represents the compression bitrate—lower values correspond to smaller compressed audio sizes and higher compression efficiency.
            Circle size denotes model complexity: smaller sizes represent more lightweight and faster models.
        }
        \label{fig:bubble_chart}
    \end{center}
    \vskip -0.3in
\end{figure}

To tackle the above challenges, we propose \system, a \underline{L}ightweight and \underline{L}oss\underline{L}ess neural \underline{A}udio \underline{C}odec.
\system employs one quantizer for discrete token generation, eliminating the need for hierarchically structured tokens.
Our design leverages lightweight convolutional networks and local Transformer modules to enhance feature extraction and processing efficiency.
Furthermore, we introduce TConv, a novel module designed to capture both short- and long-term acoustic variations.
This module enables high-fidelity audio reconstruction while maintaining low computational overhead.

The advantages of \system are multifaceted.
By employing just one quantizer and significantly reducing computational and memory requirements, it can be easily and cost-effectively integrated into diverse audio processing applications.
Our experimental results also demonstrate that, despite its architectural simplicity, \system outperforms SOTA methods in terms of both audio quality and token generation quality.
Comprehensive evaluations across diverse datasets further validate its robustness, adaptability, and efficiency, making it a compelling choice for modern audio tasks.

We summarize the key contributions of this work as follows:
\begin{itemize}
    \item \textbf{Efficient Model Architecture:} The proposed design achieves reduced parameter count and computational cost by an order of magnitude while maintaining high performance.
    \item \textbf{Single-Quantizer Design:} \system eliminates the need for hierarchical quantization, simplifying token generation and downstream integration.
    \item \textbf{Enhanced Feature Extraction for Audio Data:} We introduce TConv, a novel module designed to capture both short- and long-term acoustic variations, facilitating higher-fidelity audio reconstruction.
\end{itemize}

%% file: body/related_works.tex
\section{Related Works}

\subsection{Neural Audio Codecs}
Traditional audio codecs \cite{valin2012, dietz2015, neuendorf2013} mainly rely on signal processing techniques such as linear predictive coding and transform coding for compressing audio signals.
While effective, these methods often depend on manually engineered designs, which can limit their flexibility and applicability.
The emergence of neural audio codecs has introduced a paradigm shift by adopting data-driven frameworks that learn efficient audio representations from large datasets.
These neural models \cite{SoundStream2021, HiFi-Codec2023X, encodec2023, APCodec2024, dac2024,FunCodec2024, LightCodec2024, Single-codec2024} typically employ an encoder-decoder architecture.
The encoder compresses input audio into a compact latent feature and then quantizes it into a discrete representation, while the decoder reconstructs the audio from this representation.
A significant milestone in this domain was achieved by SoundStream \cite{SoundStream2021}, which introduced a fully convolutional encoder-decoder network integrated with a residual vector quantizer (RVQ). This innovation enabled the unified handling of diverse audio types, such as speech and music, demonstrating robust performance across various domains.

\subsection{High-Fidelity Multi-Quantizer Audio Codecs}
Recent advancements in audio codec have increasingly focused on multi-quantizer architectures \cite{SoundStream2021, HiFi-Codec2023X, FunCodec2024, encodec2023, dac2024} to improve reconstruction fidelity and minimize compression errors.
EnCodec \cite{encodec2023} pushed the boundaries of performance by employing a sophisticated network architecture and introducing a novel loss design.
Building on these efforts, DAC \cite{dac2024} introduced further innovations, including the Snake activation function, the improved RVQ, and a larger network design.
In addition, DAC optimized both adversarial and reconstruction loss functions, achieving SOTA performance in audio compression.
However, despite their impressive results, these models are computationally intensive and challenging for integration into other types of downstream tasks due to their multi-quantizer architecture, which often limits their applicability in real-world scenarios.

\subsection{Single-Quantizer Audio Codecs}

Single-quantizer audio codecs have recently garnered significant interest for their ability to produce a single, unified stream of discrete tokens, simplifying integration with downstream generative models.
However, many prominent models are optimized for high-fidelity reconstruction within specific domains.
For instance, SingleCodec~\cite{Single-codec2024} and TAAE~\cite{TAAE2025} demonstrate strong performance on speech, but their generalization to diverse, out-of-domain audio remains a challenge.
To address this limitation, approaches like WavTokenizer~\cite{WavTokenizer2025} and UniCodec \cite{UniCodec2025} have been developed for general-purpose audio types.
However, this broader scope often comes at the cost of reconstruction accuracy, with objective metrics such as PESQ~\cite{pesq2001} and STOI~\cite{stoi2010} falling short.
Furthermore, these models tend to be computationally intensive, limiting their practical applicability.
% ! 复杂的训练策略，庞大的参数模型

Consequently, a clear research gap exists for a single-quantizer codec that simultaneously achieves high objective fidelity, cross-domain generalizability, and computational efficiency.
Our proposed model, \system, is designed to fill this void by introducing a lightweight, streamable architecture that excels in high-quality reconstruction across varied audio types.

%% file: body/model.tex
\section{L3AC}

\system is a lightweight and efficient neural audio codec that delivers high-fidelity audio with low computational complexity and strong scalability.
This section outlines the design components of \system, beginning with a discussion of its core structure aimed at achieving a sufficient receptive field, followed by a detailed overview of the complete model.

\begin{figure}[t]
    \begin{center}
        \centerline{\includegraphics[width=\columnwidth]{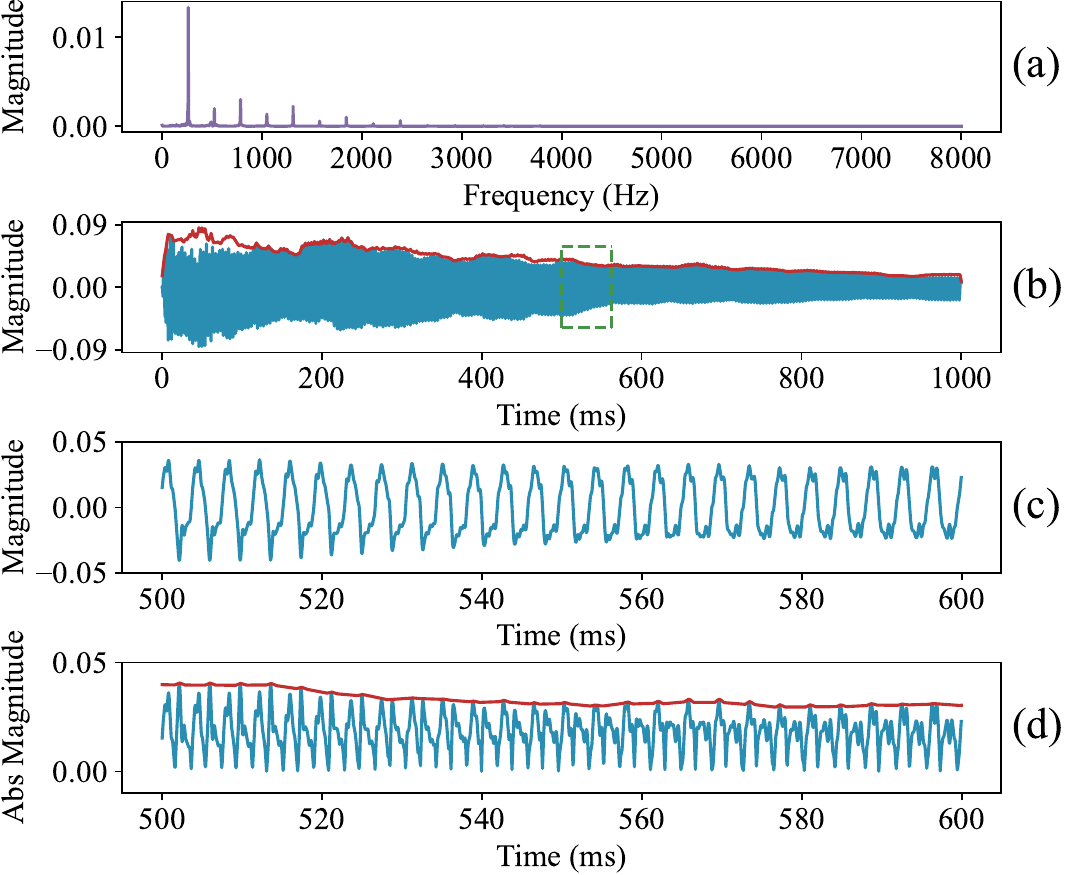}}
        \caption{
            Visualization illustrating the impact of \textbf{TPooling} on
            audio signals, showcasing its effectiveness in processing and capturing long-term signal-level features.
        }
        \label{fig:trend_pool}
        \vskip -0.35in
    \end{center}
\end{figure}

\subsection{Efficient Receptive Field Expansion}

One of the key challenges in designing effective neural audio codecs lies in achieving a sufficiently large receptive field to capture both fine-grained details and long-range signal dependencies.
This capacity is critical for preserving audio quality, as demonstrated by our ablation study.

Conventional approaches such as DAC \cite{dac2024} and EnCodec \cite{encodec2023} rely on deep convolutional stacks to enlarge the receptive field, which introduces considerable computational overhead during training and inference.
Other solutions based on Recurrent Neural Network (RNN) or transformer architectures \cite{Single-codec2024, TAAE2025} effectively capture global context but still suffer from limited parallelization or substantial computational demands.
To address these limitations, we propose a hybrid architecture that combines convolutional layers with local transformers.
This design achieves comparable receptive fields with fewer layers, enabling lower latency and reduced complexity, while remaining suitable for streaming scenarios.

Critically, receptive field expansion should be considered at both the \textbf{acoustic} and \textbf{signal} levels.
As shown in \cref{fig:trend_pool}, we analyze a one-second audio clip from a piano\footnote{\tiny https://en.wikipedia.org/wiki/C\_(musical\_note)}.
This audio primarily consists of a fundamental frequency and its overtones, producing distinct time-domain signal variations shown in \Cref{fig:trend_pool}(b).
Zooming into a specific segment, highlighted by the green dotted square in \cref{fig:trend_pool}(c), reveals both short-term variations (at the sampling-point scale) and long-term trends (approximately 10 milliseconds).
While the hybrid network allows deeper layers to extract high-level acoustic representations, shallow convolutional layers still struggle to model longer signal-level patterns, \eg, the 10ms trends mentioned above.
Dilated convolutions \cite{holschneider1990, shensa1992, yu2016X} have been explored to address this, but their effectiveness is often diminished by the periodic nature of audio signals, where dominant short-term variations overshadow the long-term trends.

To bridge this gap, we introduce \textbf{TPooling}, a novel pooling structure designed to explicitly capture global amplitude variations.
Formally, TPooling is defined as:
\begin{equation}
    \label{eq:t_pooling}
    % \begin{aligned}
    % &
    TPooling(x,K) = AvgP(MaxP(|x|,K),K),
    % \end{aligned}
\end{equation}
where $|x|$ represents the absolute value of the input signal, $K$ denotes the kernel size, and AvgP and MaxP refer to average and maximum pooling operations, respectively.

As shown in \cref{fig:trend_pool}(d), the blue line represents $|x|$, while the red line shows the result of $TPooling(x, K)$.
Similarly, in \cref{fig:trend_pool}(b), the red line illustrates the TPooling output over the full audio signal.
It can be seen that this two-stage pooling process effectively smooths over short-term fluctuations while preserving longer-term patterns;
therefore, we propose incorporating TPooling with other network modules to enhance its ability to capture long-term signal dynamics essential for high-quality audio reconstruction.

\begin{figure*}[t]
    \vskip 0.2in
    \begin{center}
        \centerline{\includegraphics[width=.8\textwidth]{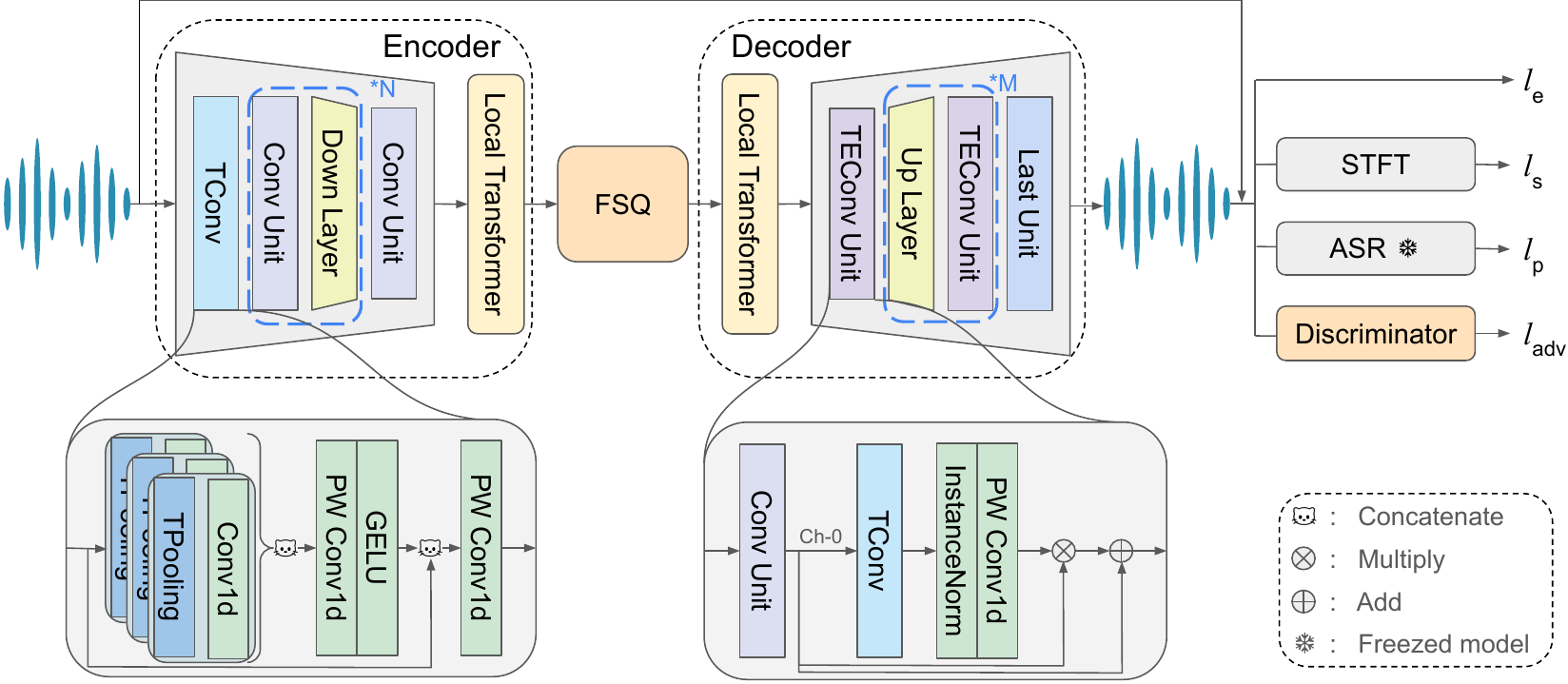}}
        \caption{
            Overview of the \system architecture, comprising an encoder, a quantizer (using Finite Scalar Quantization, FSQ), a decoder, and a discriminator. The model is optimized using reconstruction losses ($l_e$, $l_s$), perceptual loss ($l_p$), and adversarial loss ($l_{adv}$).
        }
        \label{fig:model}
    \end{center}
    \vspace{-0.25in}
\end{figure*}

\subsection{Model Design}\label{subsec:model}

The architecture of \system consists of four main components: an encoder, a quantizer, a decoder, and a discriminator.
Each is tailored to balance fidelity, efficiency, and adaptability, ensuring suitability for a wide range of deployment scenarios.

\subsubsection{Encoder.}

The encoder converts raw audio waveforms into compact, informative representations that capture temporal patterns across multiple scales.
The encoding process begins with the \textit{TConv Unit}, which is designed to establish a large receptive field at the signal level from the outset.
It first applies \textit{TPooling} with varying kernel sizes, followed by a convolutional layer to capture temporal variations at different resolutions.
These multi-scale features are concatenated and processed by a point-wise convolution that expands the channel dimension fourfold.
The result is passed through a GELU activation function to generate latent representations.
These are then concatenated with the original input and compressed back to the original channel size via another point-wise convolution, producing the final TConv output.

Next, the output is passed through a series of \textit{Conv Units} and \textit{Down Layers}.
The Conv Unit is an adaptation of the ConvNeXt architecture \cite{ConvNext2022} for time-domain audio processing.
Here, 2D convolutions are replaced with 1D convolutions.
Additionally, the Snake activation function \cite{snake2020} is employed to effectively capture the periodic and nonlinear characteristics of audio, as demonstrated in prior studies \cite{dac2024, BigVGAN2022}.
This design achieves a balance between computational efficiency and high-fidelity feature extraction.
Following the Conv Unit, the Down Layer applies a strided convolutional network to perform downsampling, a method commonly adopted in previous studies \cite{encodec2023, dac2024, HiFi-Codec2023X}.
This reduces the data resolution while preserving critical information, improving processing efficiency in subsequent stages.

Finally, the downsampled features are fed into a \textit{Local Transformer}, which provides a large acoustic-level receptive field with low computational overhead and latency.
To maintain causality and ensure real-time performance, the transformer avoids backward dependencies.
Its self-attention mechanism dynamically prioritizes relevant segments of the input, yielding high-quality acoustic representations.
Notably, for low-bitrate models, an additional downsampling layer is added between the Local Transformer layers to further compress the audio.

\subsubsection{Quantizer.}

% At the core of \system, the quantizer employs a single Finite Scalar Quantization (FSQ) \cite{FSQ2023} to discretize features extracted by the encoder.
% This streamlined quantization design reduces computational overhead while preserving performance.
At the core of \system is a single quantizer used to discretize the features extracted by the encoder.
This design choice simplifies the model, reduces resource consumption, and enables seamless integration into downstream tasks.
However, single-quantizer architectures inherently produce smaller codebooks compared to multi-quantizers, which can significantly degrade model accuracy \cite{dac2024}.
Additionally, traditional single quantizers often fail when codebooks become large, leading to collapse.
To mitigate this, we adopt Finite Scalar Quantization (FSQ) \cite{FSQ2023}, which supports large codebooks without risking collapse issues.
FSQ uses straightforward scalar quantization in a bounded low-dimensional space, reducing complexity, providing faster quantization speeds and greater stability during training.

Additionally, inspired by \cite{TAAE2025}, we adopt a hybrid quantization strategy.
With a 50\% probability, features output by the encoder are perturbed with uniform noise \cite{brendel2024} rather than being quantized.  % improving robustness during training.

\subsubsection{Decoder.}

The decoder reconstructs high-quality audio from quantized representations.
While it mirrors the encoder's architecture, it incorporates several enhancements to improve reconstruction fidelity.

The decoding process begins with a Local Transformer, which processes the quantized features to restore long-range contextual information.
This helps reestablish the acoustic coherence lost during compression.
Next, the features are passed through a \textit{TEConv Unit}, which extends the encoder's Conv Unit by introducing temporal attention.
Specifically, it applies a TConv layer along the temporal axis to extract multi-scale latent representations.
These are normalized using InstanceNorm and passed through a point-wise convolution to generate temporal attention weights.
The original Conv Unit output is then modulated using these weights via an attention-like mechanism, improving the temporal accuracy and perceptual quality of the reconstruction.

Following temporal enhancement, the features undergo resolution restoration through the \textit{Up Layer}.
Instead of using transposed convolutions—which are prone to introducing artifacts—this layer adopts linear upsampling to preserve audio integrity while efficiently recovering the original resolution.
Finally, the \textit{Last Unit} completes the reconstruction.
This component consists of deeper convolutional layers with increased parameter capacity and Snake activation functions.
The additional parameters allow the model to recover fine-grained details, further enhancing the naturalness and overall quality of the reconstructed audio.

\subsubsection{Discriminator.}

To encourage realistic audio generation, \system incorporates the multi-scale discriminator architecture used in DAC \cite{dac2024}.
This discriminator operates on both time-domain and frequency-domain signals, thereby providing stronger gradient feedback to \system's generator (\ie, the encoder and decoder) and helping mitigate periodicity artifacts, as proved in prior work \cite{UnivNet2021, BigVGAN2022}.
To ensure balanced training, the discriminator is updated less frequently (\eg, every 15–25 steps), preventing it from converging faster than the generator.

\input{tab/table_cmd}

\begin{table*}[ht]
    % \vskip 0.15in
    \begin{center}
        \vskip -0.15in
        \input{tab/signal_evaluation}
    \end{center}
    \vskip -0.15in
    \caption{
        \textbf{Signal-level evaluation results} for various codec models.
        \textbf{\#Q} refers to the number of quantizers used in the model.
        \textbf{MACs} (Multiply-Accumulate Computations) indicate the total number of
        arithmetic operations required to process 10 seconds of audio.
        $\dag$ denotes that the model is trained specifically for speech.
    }
    \label{tab:signal_evaluation}
\end{table*}

\subsection{Training Strategy}

\system is optimized using a combination of four loss functions (formally defined in \appendixref{A}) to enable lossless audio reconstruction:
(1) \textbf{Element-Wise Loss ($l_e$)} evaluates time-domain signal reconstruction by comparing the generated audio with the input audio.
(2) \textbf{Spectrogram Loss ($l_s$)} ensures fidelity across multiple frequency scales.
(3) \textbf{Perception Loss ($l_p$)} compares intermediate features from an automatic speech recognition (ASR) model (e.g., Whisper-tiny \cite{Whisper2023}) between original and reconstructed audio, preserving perceptual quality.
(4) \textbf{Adversarial Loss ($l_{adv}$)} is calculated by the discriminator, which guarantees realistic audio generation.

The influence of these losses evolves over training.
Early stages prioritize $l_e$ and $l_s$, while later stages emphasize $l_p$ and $l_{adv}$ to refine perceptual quality.
However, all four losses tend to decrease simultaneously.
To manage the varying impact of each loss while avoiding manual adjustment of weights at different training stages,
we employ a loss weight clamping mechanism (see \cref{eq:clamp_max}).
\begin{equation}
    \label{eq:clamp_max}
    clamp(l, max) =
    \begin{cases}
        l                               , & l < max \\
        \dfrac{l \times max}{l.detach()} , & l \geq max
    \end{cases}
\end{equation}
This approach limits the maximum value of certain losses (\eg, $l_p$, $l_{adv}$) in the early stages, while allowing other losses (\eg, $l_e$, $l_s$) to become more influential, and these effects are reversed once the network enters later stages of training, \ie, when the reconstructed audio quality has improved to a certain extent.
Additionally, we adopt the One Cycle Learning Rate policy \cite{OneCycleLR2019} to dynamically adjust the learning rate, accelerating convergence and improving final performance.

%% file: tab/table_cmd.tex
\newcommand{\da}{$\downarrow$}
\newcommand{\ua}{$\uparrow$}
\newcommand{\tabF}[1]{$^{\mathrm{#1}}$}
\newcommand{\stack}[2]{\makecell{#1\\#2}}
\newcommand{\mbps}{\stack{Bitrate}{(kbps)}}
\newcommand{\mfps}{\stack{Frame rate}{(fps)}}
\newcommand{\mmacs}{\stack{MACs\da}{(G)}}
\newcommand{\mpara}{\stack{\#Params\da}{(M)}}
\newcommand{\pu}[1]{\stack{#1}{(\%)}}
\newcommand{\fv}[1]{\textbf{#1}}
\newcommand{\sv}[1]{\underline{#1}}
\definecolor{SQColor}{rgb}{0.85,0.85,0.85}

%% file: tab/signal_evaluation.tex
\begin{tabular}{lccccccccccc}
    \toprule
    Model            & \mbps & \mfps & \#Q & \mmacs    & SDR\ua     & MEL\da    & STOI\ua   & PESQ\ua    & \stack{Audio}{SDR\ua} & \stack{Audio}{MEL\da} \\
    \midrule
    DAC              & 0.50  & 50    & 1   & 556.00    & -7.93      & 1.95      & 0.58      & 1.08       & -10.31                & 2.57                  \\
    TAAE\dag         & 0.39  & 25    & 1   & 374.89    & 0.86       & 1.51      & 0.76      & 1.53       & -7.86                 & 2.45                  \\
    TAAE\dag         & 0.70  & 25    & 2   & 374.89    & \fv{2.83}  & 1.45      & \sv{0.79} & \sv{1.64}  & \sv{-7.02}            & 2.42                  \\
    WavTokenizer     & 0.48  & 40    & 1   & 34.26     & -2.80      & \sv{1.14} & 0.75      & 1.51       & -11.37                & \sv{1.44}             \\
    WavTokenizer\dag & 0.9   & 75    & 1   & 64.17     & -1.31      & \fv{0.97} & 0.77      & 1.58       & -12.29                & \fv{1.42}             \\
    \rowcolor{SQColor}
    \textbf{\system} & 0.75  & 44    & 1   & \fv{1.55} & \sv{1.83}  & \sv{1.14} & \fv{0.82} & \fv{ 1.68} & \fv{-2.16}            & 1.89             \\
    \midrule
    DAC              & 1.0   & 50    & 2   & 556.01    & -4.99      & 1.31      & 0.70      & 1.18       & \sv{-7.47}            & 2.05                  \\
    UniCodec         & 1.05  & 75    & 1   & 71.22     & \sv{-0.70} & \fv{0.86} & \fv{0.84} & \fv{1.91}  & -8.24                 & \fv{1.26}             \\
    \rowcolor{SQColor}
    \textbf{\system} & 1.0   & 59    & 1   & \fv{2.03} & \fv{2.88}  & 1.09      & \fv{0.84} & \sv{1.77}  & \fv{-0.99}            & 1.86                  \\
    \midrule
    Encodec          & 1.5   & 75    & 2   & 55.95     & \sv{1.47}  & 1.30      & \sv{0.80} & \sv{1.50}  & \sv{-0.58}            & \fv{1.51}             \\
    DAC              & 1.5   & 50    & 3   & 556.02    & -3.23      & \sv{1.11} & 0.76      & 1.31       & -5.36                 & 1.89                  \\
    \rowcolor{SQColor}
    \textbf{\system} & 1.5   & 89    & 1   & \fv{2.37} & \fv{4.34}  & \fv{1.02} & \fv{0.86} & \fv{1.91}  & \fv{0.92}             & \sv{1.83}             \\
    \midrule
    Encodec          & 3.0   & 75    & 4   & 55.95     & \sv{4.42}  & 1.15      & 0.86      & 1.91       & \sv{1.78}             & \fv{1.41}             \\
    DAC              & 3.0   & 50    & 6   & 556.05    & -0.77      & \fv{0.83} & \sv{0.87} & \sv{1.92}  & -1.43                 & \sv{1.67}             \\
    \rowcolor{SQColor}
    \textbf{\system} & 3.0   & 167   & 1   & \fv{1.64} & \fv{6.75}  & \sv{0.92} & \fv{0.90} & \fv{2.31}  & \fv{3.89}             & 1.74                  \\
    \bottomrule
\end{tabular}

%% file: body/experiments.tex
\begin{table*}[ht]
    \begin{center}
        \vskip -0.15in
        \input{tab/application_evaluation}
    \end{center}
    \vskip -0.12in
    \caption{
        \textbf{Application-level evaluation results} for various codec models.
        \textbf{ASR} represents the automatic speech recognition task,
        \textbf{ASV} represents the automatic speaker verification task,
        \textbf{ER} represents the emotion recognition task,
        and \textbf{AEC} represents the audio event classification task.
    }
    \label{tab:application_evaluation}
\end{table*}

\section{Experiments and Results}

\subsection{Experimental Settings}\label{sec:Experimental_Settings}

A summary of our experimental setup is provided below.
Additional implementation details can be found in \appendixref{B}.

\subsubsection{Training Datasets.}
\system was trained using datasets from three audio domains:
1) \textbf{Speech Domain:} The “train-clean-100” and “train-clean-360” subsets of LibriSpeech \cite{Librispeech2015} for clean speech, alongside the “cv-corpus-18.0” dataset from Common Voice \cite{commonvoice} for noisy speech.
2) \textbf{Music Domain:} The low-quality version of MTG-Jamendo dataset \cite{mtg2019}.
3) \textbf{General Audio Domain:} The FSD50K dataset \cite{FSD50K2022}, which includes a wide range of audio categories.

Compared to other works, \system was trained on a relatively smaller collection of datasets, as detailed in \appendixref{B}.
All audio was uniformly resampled to 16,000 Hz, as our empirical results indicated that higher sampling rates did not yield performance gains and substantially increased the computational load.
During training, batches were alternated between these datasets, with samples randomly selected from each dataset.
This approach ensured consistent training time and equal weight for each dataset, facilitating balanced learning across different audio domains.

\subsubsection{Training Details.}
We optimized \system using the AdamW optimizer, incorporating a one-cycle learning rate schedule \cite{OneCycleLR2019}.
The learning rate warmed up from $5 \times 10^{-5}$ to a peak of $5 \times 10^{-4}$ before decaying to $5 \times 10^{-6}$.
Gradient clipping was employed to stabilize training, with maximum norms of $10,000$ for the codec and $10$ for the discriminator.
Additionally, a weight decay of $1 \times 10^{-5}$ was applied during discriminator training.
The training was conducted on a single NVIDIA RTX 4090 GPU, demonstrating the model's computational efficiency.

\subsubsection{Evaluation Details.}
Evaluation is conducted from two aspects, \ie, assessing the quality of both the reconstructed audio and the discrete tokens generated by the codec.

\textbf{\textit{Reconstructed Audio Evaluation.}}
This evaluation was conducted using the Codec-SUPERB benchmark \cite{CodecSuperB2024}, which provides both signal-level and application-level assessments for the reconstructed audio.
Specifically, the datasets and evaluation metrics were sourced from the Codec-SUPERB challenge held at SLT 2024\footnote{\tiny https://github.com/voidful/Codec-SUPERB/tree/SLT\_Challenge}.

1) Signal-Level Evaluations:
We measured Signal-to-Distortion Ratio (SDR) \cite{mir_eval2014}, Perceptual Evaluation of Speech Quality (PESQ) \cite{pesq2001}, Short-Time Objective Intelligibility (STOI) \cite{stoi2010}, and Mel Spectrogram Distance (MEL) across 11 datasets covering speech, music, and general audio.

2) Application-Level Evaluations:
We incroporate four downstream tasks: automatic speech recognition (ASR) on the LibriSpeech dataset, automatic speaker verification (ASV) on the VoxCeleb dataset \cite{Voxceleb2020}, emotion recognition (ER) on the RAVDESS dataset \cite{RAVDESS2019}, and audio event classification (AEC) on the ESC-50 dataset \cite{ESC2015}.

\textbf{\textit{Generated Tokens Evaluation.}}
Inspired by DASB \cite{DASB2024X}, we evaluated the quality of the generated tokens on two representative downstream tasks: Automatic Speech Recognition (ASR) with tokens generated by codec models and Text-to-Speech (TTS) using tokens to synthesize audio.
For ASR, we trained a transcription model on the LibriSpeech dataset using an LSTM-based architecture from SpeechBrain \cite{speechbrain2021X}.
For TTS, we trained a synthesis model on the LJSpeech dataset \cite{ljspeech17} using a Transformer-based architecture from SpeechBrain.
All models were trained under identical settings (including same hyper-parameters, same software versions and same hardware, \etc.) to support fair comparison.

\textbf{\textit{Baselines.}}
We compared \system against several SOTA models: the multi-quantizer codecs EnCodec \cite{encodec2023} and DAC \cite{dac2024}, and the single-quantizer models WavTokenizer \cite{WavTokenizer2025}, TAAE \cite{TAAE2025}, and UniCodec \cite{UniCodec2025}.
For EnCodec and DAC, performance was evaluated across multiple bitrates by adjusting the number of RVQ levels.
For WavTokenizer, the “large-600-24k-4096” and “large-320-24k-4096” pretrained models were used, as they represent the most extensively trained versions of the model.
For TAAE and UniCodec, official pretrained checkpoints were used.

\begin{table*}[t]
    \begin{center}
      \setlength{\tabcolsep}{4pt} %4pt
        \vskip -0.15in
        \input{tab/downstream_evaluation}
    \end{center}
    \vskip -0.15in
    \caption{
        \textbf{Generated Tokens Evaluation} on various codec models.
        \textbf{ASR-1} denotes the ASR task on the LibriSpeech test-clean subset,
        \textbf{ASR-2} is on the test-other subset,
        \textbf{TTS} represents the text-to-speech task,
        and \textbf{WER(Total)} is the mean of the Word Error Rates from all three tasks.
    }
    \label{tab:downstream_evaluation}
\end{table*}

\subsection{Results}

\subsubsection{Reconstructed Audio Quality.}

We evaluated \system's reconstruction performance against SOTA baseline models across various bitrates, with detailed results presented in \cref{tab:signal_evaluation} (signal-level metrics) and \cref{tab:application_evaluation} (application-level metrics).
The findings show that \system consistently matches or exceeds the performance of SOTA codecs while operating with significantly lower computational complexity.
For a more intuitive comparison, \cref{fig:bubble_chart} plots STOI scores against model bitrate and model complexity, highlighting \system's exceptional efficiency and effectiveness.
Since reconstruction quality is directly tied to the compression rate (\ie, bitrate), we partition our analysis into three distinct bitrate ranges to evaluate \system's performance in the following.

\textbf{\textit{Performance below 1 kbps.}}
In the challenging sub-1 kbps range, \system demonstrates a decisive advantage.
On application-level tasks, \system operating at just 0.75 kbps achieves an ASR WER of 5.01\% and an AEC accuracy of 65.0\%, outperforming all competing models (\cref{tab:application_evaluation}).
Its superiority is reinforced by leading signal-level metrics, including an STOI of 0.82, a PESQ of 1.68, and an Audio SDR of -2.16 (\cref{tab:signal_evaluation}).
While its Speech/Audio MEL distance is marginally higher than WavTokenizer and its Speech SDR is slightly behind the speech-specialized TAAE model, \system delivers the strongest and most versatile overall performance, excelling across a wide range of tasks while operating at a fraction of the computational cost.

\textbf{\textit{Performance at 1 kbps.}}
At 1 kbps, \system remains highly competitive with the top-performing models.
Although UniCodec achieves a better PESQ score and MEL distance, \system achieves a significantly better SDR (2.88 vs. -0.70) and Audio SDR (-0.99 vs. -8.24).
Furthermore, while \system's application-level performance is comparable to UniCodec's, its efficiency is orders of magnitude greater.
UniCodec's computational demand of 71.22G MACs is over 35 times higher than \system's 2.03G MACs, establishing \system as a far more practical and efficient solution for achieving high-fidelity results.

\textbf{\textit{Performance above 1 kbps.}}
In this bitrate range, we compare \system against multi-quantizer models like Encodec and DAC.
The results demonstrate the superiority of \system's single-quantizer architecture, as it outperforms these more complex models on nearly every metric.
At the signal level, \system establishes a commanding lead in PESQ scores; for instance, at 3.0 kbps, its PESQ of 2.31 far exceeds Encodec's 1.91 and DAC's 1.92.
This trend continues in application-level evaluations, where \system is significantly better than Encodec and DAC across all settings, with only one trivial exception: its AEC accuracy at 3 kbps (80.9\%) is negligibly lower than Encodec's (81.0\%).
This proves that \system's design is fundamentally more efficient, capable of reconstructing lossless audio in a single stream of discrete units where other models require multiple, complex layers of quantization.

Across all benchmarks, MEL distance is the only metric where \system does not consistently lead.
This is an expected outcome of our model's design.
While competing models directly optimize using Mel spectrum-based loss, \system employs an STFT-based loss function that prioritizes a broader set of spectral features.
% contributing to its strong performance on other critical metrics.

\subsubsection{Generated Tokens Quality.}

Beyond reconstruction, we evaluated the quality of \system's generated tokens for downstream tasks, including automatic speech recognition (ASR) and text-to-speech (TTS), with results shown in Table \ref{tab:downstream_evaluation}.

It is important to note that the TAAE model was specifically trained on the LibriSpeech dataset, which is used for the ASR-1 and ASR-2 evaluations.
Unsurprisingly, it performs well on these tasks but fails catastrophically on the TTS task which uses the out-of-domain LJSpeech dataset (WER $> 97\%$), revealing its poor generalization.

Excluding the TAAE model, \system leads in ASR and remains highly competitive in TTS.
At 1.0 kbps, it achieves a TTS WER of 27.18\%, narrowly trailing WavTokenizer (26.20\%) but outperforming other models, \eg, UniCodec (35.69\%), by a significant margin.
This well-rounded performance indicates that \system’s tokens are semantically rich and generalize effectively across different tasks and datasets.

Our analysis also confirms that increasing the number of quantizers in token-based codecs degrades their usability in downstream applications, \eg, TTS task, reaffirming the architectural limitations of such models.
\system's single-quantizer approach, by contrast, produces a simplified token sequence that enhances compatibility with standard downstream models.

Further analysis reveals a trade-off between frame rate and bitrate.
Lower frame rates reduce sequence length, easing the burden on downstream models, but also limit information density due to a lower bitrate if the codebook size is fixed.
Our experiments suggest that a bitrate of around 1 kbps with a 60 fps frame rate offers an effective compromise, maximizing utility while maintaining manageable sequence lengths.

\subsubsection{Ablation study.}

Our ablation studies demonstrate the critical role of each design choice.
In particular, shrinking the Local Transformer window or replacing TConv with a standard convolution causes severe performance drops or even convergence failures, underscoring the central importance of a sufficiently large receptive field—an aspect seldom emphasized in prior work.
Likewise, omitting the perceptual loss term or disabling our loss weight clamp mechanism during training yields a marked degradation across evaluation metrics, validating the necessity of our loss design.
Together, these results confirm that both our architectural innovations and training designs are indispensable for achieving robust, high-quality audio modeling.
More details can be found in \appendixref{C}.

%% file: tab/application_evaluation.tex
\begin{tabular}{lcccccccc}
    \toprule
    Model            & \mbps & \mfps & \#Q & \mmacs    & \pu{WER(ASR)\da} & \pu{EER(ASV)\da} & \pu{Acc(ER)\ua} & \pu{Acc(AEC)\ua} \\
    \midrule
    DAC              & 0.50  & 50    & 1   & 556.00    & 78.12            & 34.54            & 22.50           & 13.90            \\
    TAAE\dag         & 0.39  & 25    & 1   & 374.89    & 12.12            & 15.20            & 57.43           & 31.70            \\
    TAAE\dag         & 0.70  & 25    & 2   & 374.89    & \sv{9.98}        & 12.49            & \sv{61.88}      & 34.50             \\
    WavTokenizer     & 0.48  & 40    & 1   & 34.26     & 16.65            & \sv{12.10}        & 55.07           & \sv{42.65}       \\
    WavTokenizer\dag & 0.9   & 75    & 1   & 64.17     & 13.21            & 12.31            & 58.68           & 41.55            \\
    \rowcolor{SQColor}
    \textbf{\system} & 0.75  & 44    & 1   & \fv{1.55} & \fv{5.01}        & \fv{9.52}        & \fv{65.49}      & \fv{65.00}       \\
    \midrule
    DAC              & 1.0   & 50    & 2   & 556.01    & 20.65            & 18.94            & 40.28           & 26.70             \\
    UniCodec         & 1.05  & 75    & 1   & 71.22     & \sv{6.65}        & \fv{5.14}        & \fv{70.90}       & \sv{58.40}        \\
    \rowcolor{SQColor}
    \textbf{\system} & 1.0   & 59    & 1   & \fv{2.03} & \fv{4.51}        & \sv{8.03}        & \sv{68.75}      & \fv{69.65}       \\
    \midrule
    Encodec          & 1.5   & 75    & 2   & 55.95     & 11.19            & \sv{10.72}       & \sv{57.57}      & \sv{62.70}        \\
    DAC              & 1.5   & 50    & 3   & 556.02    & \sv{9.66}        & 10.99            & 50.35           & 40.85            \\
    \rowcolor{SQColor}
    \textbf{\system} & 1.5   & 89    & 1   & \fv{2.37} & \fv{4.11}        & \fv{6.42}        & \fv{70.42}      & \fv{72.90}       \\
    \midrule
    Encodec          & 3.0   & 75    & 4   & 55.95     & 4.59             & 4.32             & 68.12           & \fv{81.00}          \\
    DAC              & 3.0   & 50    & 6   & 556.05    & \sv{4.26}        & \sv{3.56}        & \sv{70.00}       & 70.40             \\
    \rowcolor{SQColor}
    \textbf{\system} & 3.0   & 167   & 1   & \fv{1.64} & \fv{3.41}        & \fv{3.46}        & \fv{71.81}      & \sv{80.90}        \\
    \bottomrule
\end{tabular}

%% file: tab/downstream_evaluation.tex
\begin{tabular}{lcccccccc}
    \toprule
    Model            & \mbps & \mfps & \#Q & \mmacs & \pu{WER(ASR-1)\da} & \pu{WER(ASR-2)\da} & \pu{WER(TTS)\da} & \pu{WER(Total)\da} \\
    \midrule
    TAAE\dag         & 0.39  & 25    & 1   & 374.89 & 33.43              & 53.84              & 97.57            & 61.61              \\
    TAAE\dag         & 0.7   & 25    & 2   & 374.89 & \fv{27.87}         & \fv{49.16}         & 97.51            & 58.18              \\
    \midrule
    DAC              & 0.5   & 50    & 1   & 556.00 & 78.90              & 89.66              & \sv{26.93}       & 65.19              \\
    DAC              & 1.0   & 50    & 2   & 556.01 & 63.17              & 83.12              & 35.69            & 60.66              \\
    DAC              & 1.5   & 50    & 3   & 556.02 & 57.34              & 80.29              & 41.25            & 59.63              \\
    WavTokenizer     & 0.48  & 40    & 1   & 34.26  & 64.23              & 84.23              & \fv{26.20}       & 58.22              \\
    WavTokenizer\dag & 0.9   & 75    & 1   & 64.17  & 59.51              & 82.47              & 36.48            & 59.49              \\
    UniCodec         & 1.05  & 75    & 1   & 71.22  & 49.35              & 75.86              & 35.03            & 53.41              \\
    \rowcolor{SQColor}
    \textbf{\system} & 0.75  & 44    & 1   & 1.55   & 41.41              & 68.70              & 27.83            & 45.98              \\
    \rowcolor{SQColor}
    \textbf{\system} & 1.0   & 59    & 1   & 2.03   & \fv{38.61}         & \fv{66.76}         & 27.18            & \fv{44.18}         \\
    \rowcolor{SQColor}
    \textbf{\system} & 1.5   & 89    & 1   & 2.37   & \sv{39.55}         & \sv{68.40}         & 27.98            & \sv{45.31}         \\
    \bottomrule
\end{tabular}

%% file: body/conclusion.tex
\section{Conclusion}

This work introduces \system, a lightweight and lossless audio codec, 
% a significant advancement in neural audio compression.
% by employing a single-quantizer architecture, 
which effectively addresses fundamental challenges related to scalability, adaptability, and computational efficiency.
Our comprehensive experimental results demonstrate that \system achieves audio quality on par with, and often exceeding, that of more complex SOTA systems—while operating at a fraction of their computational cost.
This inherent efficiency, coupled with a simplified training pipeline, substantially reduces the training burden and facilitates rapid adaptation for specialized audio tasks.
Consequently, \system emerges as a superior choice for researchers and developers seeking an efficient, high-quality, and adaptable neural audio compression solution.

%% file: body/others.tex

%\section*{Acknowledgements}
%
%\textbf{Do not} include acknowledgements in the initial version of
%the paper submitted for blind review.
%
%If a paper is accepted, the final camera-ready version can (and
%usually should) include acknowledgements.  Such acknowledgements
%should be placed at the end of the section, in an unnumbered section
%that does not count towards the paper page limit. Typically, this will
%include thanks to reviewers who gave useful comments, to colleagues
%who contributed to the ideas, and to funding agencies and corporate
%sponsors that provided financial support.

\newpage
% \section*{Impact Statement}

% This paper presents work whose goal is to advance the field of Machine Learning.
% There are many potential societal consequences of our work, none which we feel must be specifically highlighted here.

%\nocite{xxx}

%% file: appendix/all.tex
\appendix
\onecolumn

\input{appendix/title}
\input{appendix/A_loss_func}
\input{appendix/B_training_details}
\input{appendix/C_ablation_study}

%% file: appendix/title.tex
\begin{center}
\LARGE \textbf{L3AC: Towards a Lightweight and Lossless Audio Codec}

\LARGE \textbf{Supplementary material}
\end{center}

%% file: appendix/A_loss_func.tex
\section{A. Loss Functions}

Below is a more concise, clarified presentation of each loss definition.

$\bullet$ \textbf{Element-Wise Loss} ($l_e$):
\begin{equation}
    \label{eq:loss_ele}
    l_e = \|x-\hat{x}\|_{1}
\end{equation}
$l_e$ computes the L1 distance between the ground truth audio $x$ and the reconstructed audio $\hat{x}$.
This loss serves as a fundamental measure of waveform reconstruction accuracy, directly penalizing any sample-level deviation in the time-domain signal.
By minimizing this distance, the model is encouraged to generate an output that is numerically close to the original, ensuring a baseline level of fidelity.

$\bullet$ \textbf{Spectrogram Loss} ($l_s$):
\begin{equation}
    \label{eq:loss_spec}
    l_s = \frac{1}{|I|} \sum_{i \in I}(\|S_i(x)-S_i(\hat{x})\|_{1} + \|\log_{10}(S_i(x)^2)-log_{10}(S_i(\hat{x})^2)\|_{1})
\end{equation}
$l_s$ operates in the frequency domain, which is more aligned with human auditory perception.
It averages L1-based measures across multiple STFT scales, where $S_i$ uses a window size of $2^i$ and a hop length of $2^i/4$, and $I=(5, 6, 7, 8, 9, 10, 11)$.
The first term ensures the magnitude of spectral components is similar, while the second log-magnitude term better reflects the logarithmic nature of human loudness perception.
This multi-scale approach enables the model to capture both fine-grained temporal dynamics (with smaller windows) and precise frequency information (with larger windows).

$\bullet$ \textbf{Perception Loss} ($l_p$):
\begin{equation}
    \label{eq:loss_perception}
    l_p = \|ASR(x)-ASR(\hat{x})\|_{2}
\end{equation}
$l_p$ is defined as the L2 distance between the latent features extracted by a pre-trained ASR model ($ASR$) for $x$ and $\hat{x}$.
This loss guides the reconstruction towards perceptual realism, particularly for speech.
By minimizing the distance in this feature space, we ensure the generated audio preserves essential characteristics for intelligibility, such as phonetic content.

$\bullet$ \textbf{Adversarial Loss} ($l_{adv}$):
\begin{equation}
    \label{eq:loss_adv}
    l_{adv} = (\sum_{k=1}^{K} \| 1 - D_k(\hat{x}) \|_2)
    + 2 \times (\sum_{k=1}^{K} \sum_{l=1}^{L} \| D_k^l(x) - D_k^l(\hat{x}) \|_1)
\end{equation}
$l_{adv}$ is based on a multi-scale discriminator, consistent with the DAC.
Here, $K$denotes the number of discriminators, $D_k$ denotes the k-th discriminator.
$D_k^l$ can generate the latent features of the $l$-th layer of $D_k$.
This encourages \system to improve the naturalness of the reconstructed audio at multiple levels of abstraction.

%% file: appendix/B_training_details.tex
\begin{table*}[ht]
    \vskip 0.15in
    \begin{center}
        \setlength{\tabcolsep}{3pt} %4pt
        \begin{tabular}{lcccccccc}
            \toprule
            Model                   & \stack{Bitrate}{(kbps)} & \stack{MACs}{(G)} & \stack{\#Params}{(M)} & \stack{Encoder}{rates} & \stack{Decoder}{rates} & \stack{Transformer}{window size} & \stack{Codebook}{levels} \\
            \midrule
            \textbf{\system}        & 0.75                    & 1.55              & 11.29                 & (6, 5, 4, 3)           & (5, 4, 3, 2, 3)        & 600                              & (7, 7, 7, 7, 7, 7)       \\
            \textbf{\system}        & 1.0                     & 2.03              & 11.27                 & (6, 5, 3, 3)           & (5, 3, 3, 2, 3)        & 750                              & (7, 7, 7, 7, 7, 7)       \\
            \textbf{\system}        & 1.5                    & 2.37              & 11.25                 & (6, 5, 3, 2)           & (5, 3, 3, 2, 2)        & 600                              & (7, 7, 7, 7, 7, 7)       \\
            \textbf{\system}        & 3                      & 1.64              & 10.31                 & (6, 4, 4)              & (4, 4, 3, 2)           & 400                              & (9, 9, 9, 7, 7, 7)       \\
            \midrule
            \textbf{Ablation study} & 0.75                    & 1.53$\sim$1.55    & 11.29                 & (6, 5, 4)              & (5, 4, 3, 2)           & 75$\sim$300                           & (7, 7, 7, 7, 7, 7)      \\
            \bottomrule
        \end{tabular}
    \end{center}
    \vskip -0.1in

    \caption{
        Overview of the training configurations.
        \textbf{Encoder rates} represent the downsampling factors for each Down Layer in the Encoder, while \textit{Decoder rates} denote the corresponding upsampling factors for each Up Layer in the Decoder.
        \textbf{Transformer window size} denotes the window size used by the Local Transformer.
        \textbf{Codebook levels} refers to the number of levels within the FSQ Codebook.
    }
    \label{tab:training_details}
\end{table*}

\section{B. Training Details}

As discussed in Section \ref{sec:Experimental_Settings}, we implement multiple \system variants targeting different compression rates (\ie, bitrates).
\cref{tab:training_details} summarizes the training configurations used for each variant.
These settings jointly determine each model's complexity and the resulting bitrate.
For instance, to achieve the 0.75 kbps bitrate, we assume a standard input sampling rate of 16 kHz.
The encoder uses downsampling factors of (6, 5, 4, 3), resulting in a total downsampling factor of $6 \times 5 \times 4 \times 3 = 360$.
This produces a frame rate of $16000~\text{Hz} / 360 = 44.44~\text{frames}/s$.
Each frame is then quantized using a Finite Scalar Quantization (FSQ) codebook with 6 levels, each of size 7.
The number of bits required per frame is thus $6 \times log2(7) \approx 16.84 \text{bits}$.
The final bitrate is calculated as $ 44.44~\text{frames}/s \times 16.84~\text{bits}/\text{frame} \approx 748~\text{bps} $, or 0.75 kbps.

In addition, \cref{tab:training_datasets} provides an overview of the datasets used to train the various competing methods.
A key observation is the scale of data used by leading models. DAC and UniCodec leverage the most diverse and extensive data sources, followed closely by WavTokenizer.
In contrast, our proposed \system is trained on the fewest datasets, with the only exception being TAAE, which is trained exclusively on clean speech.
Despite this limited training data, \system achieves SOTA performance, as shown in the last row of the table. The relative-STOI metric represents the ratio of a baseline model's STOI to that of \system, with values derived from Table \ref{tab:signal_evaluation} across all bitrate variants. For each method, we select its best STOI score and compare it to the corresponding STOI score from the same-bitrate variant of \system.
The results highlight the architectural efficiency and effectiveness of \system.

\begin{table*}[ht]
    \vskip 0.15in
    \begin{center}
        \begin{tabular}{lccccccc}
            \toprule
            Dataset                & \textbf{\system} & Encodec    & DAC        & WavTokenizer & UniCodec   & TAAE       \\
            \midrule
            Common Voice           & \checkmark       & \checkmark & \checkmark & \checkmark   & \checkmark &            \\
            LibriSpeech (LibriTTS) & \checkmark       &            &            & \checkmark   & \checkmark & \checkmark \\
            Libri-Light            &                  &            &            &              & \checkmark & \checkmark \\
            DNS Challenge          &                  & \checkmark & \checkmark &              &            &            \\
            VCTK                   &                  &            & \checkmark & \checkmark   & \checkmark &            \\
            DAPS                   &                  &            & \checkmark &              &            &            \\
            \midrule
            FSD50K                 & \checkmark       & \checkmark &            &              &            &            \\
            AudioSet               &                  & \checkmark & \checkmark & \checkmark   & \checkmark &            \\
            \midrule
            MTG-Jamendo            & \checkmark       & \checkmark & \checkmark & \checkmark   & \checkmark &            \\
            MUSDB                  &                  &            & \checkmark & \checkmark   & \checkmark &            \\
            \midrule
%            relative-STOI          & 0.0              & -0.04      & -0.03      & -0.05        & 0.0        & -0.03      \\
            relative-STOI          & 100\%            & 95.6\%     & 96.7\%     & 93.9\%       & 100\%      & 96.3\%     \\
            \bottomrule
        \end{tabular}
    \end{center}
    \vskip -0.1in

    \caption{
        Datasets used in different models' training.
        \checkmark~ indicates that the corresponding dataset was included in the training process for the model.
        \textbf{relative-STOI} refers to the relative performance in STOI compared to \system.
    }
    \label{tab:training_datasets}
\end{table*}

\begin{table}[h]
    % \vskip 0.15in
    \begin{center}
        \input{tab/ablation_study}

    \end{center}
    \vskip -0.1in
    \caption{
        Ablation study of \system's key components.
        \checkmark~ indicates the model variant that uses our proposed TConv module instead of traditional convolutions.
        \textbf{NC} represents the model cannot converge.
        Performance is evaluated using our \textbf{Perception Loss} ($l_p$), where lower values indicate better perceptual similarity,
        and \textbf{STOI}, an objective metric for speech intelligibility (higher is better).
    }
    \label{tab:ablation}
\end{table}

%% file: tab/ablation_study.tex
\begin{tabular}{lcccc}
    \toprule
    Model                 & \stack{Transformer}{window size} & TConv      & \stack{Perception}{Loss}\da & STOI\ua \\
    \midrule
    \textbf{\system}      & 300                              & \checkmark & 13.61                       & 0.842   \\
    \textbf{\system}      & 150                              & \checkmark & 14.52                       & 0.833   \\
    \textbf{\system}      & 75                               & \checkmark & NC                          & NC      \\
    \textbf{\system}      & 300                              &            & 13.94                       & 0.828   \\
    \textbf{\system}      & 150                              &            & NC                          & NC      \\
    \midrule
    \textbf{\system}      & 300                              & \checkmark & 13.61                       & 0.842   \\
    ~ w/o Clamping Loss   & 300                              & \checkmark & 11.72                       & 0.815   \\
    ~ w/o Perceptual Loss & 300                              & \checkmark & 41.35                       & 0.809   \\
    \midrule
    \textbf{\system}      & 150                              & \checkmark & 14.52                       & 0.833   \\
    ~ w/o Clamping Loss   & 150                              & \checkmark & NC                          & NC      \\
    ~ w/o Perceptual Loss & 150                              & \checkmark & 59.92                       & 0.769   \\
    \midrule\bottomrule
\end{tabular}

%% file: appendix/C_ablation_study.tex
\section{C. Ablation Study}

We conducted a series of ablation experiments to validate our key design choices, with all models trained on the Speech Domain datasets (LibriSpeech and Common Voice).
% due to computational constraints.
The results, presented in \cref{tab:ablation}, are summarized below.

\subsection{C.1 Receptive Field Size}
A sufficiently large receptive field is crucial for modeling audio, both at the raw signal level and the higher contextual level.

\textbf{Acoustic-level Receptive Field:}
The window size of the local attention module determines the acoustic-level receptive field.
As shown in \cref{tab:ablation}, progressively reducing this window size from our chosen value leads to a steady decline in performance, particularly harming perceptual metrics.
Critically, reducing the window to 75 resulted in training convergence failure, highlighting the need for a large contextual window to capture long-range dependencies in audio.

\textbf{Signal-level Receptive Field:}
The TConv layers are vital for expanding the signal-level receptive field, allowing the model to capture long-term audio variations and effectively reconstruct the high-fidelity audio.
Replacing TConv with standard convolutions, which restrict this receptive field, leads to a drop in reconstruction quality.
Furthermore, the TConv module improves model robustness; without it, the model's training becomes more brittle and fails to converge with a local attention window of 150—a window size that is stable in our full model.

\subsection{C.2 Loss Function Design}
Our training objective is carefully designed to achieve high-quality results and ensure training stability.

\textbf{Perceptual Loss:}
Unlike many existing models, \system combines a perceptual loss $l_p$ with traditional loss functions (such as element-wise loss $l_e$, spectrogram loss $l_s$, and adversarial loss $l_{adv}$).
This additional constraint aims to enhance the perceptual quality of the generated audio, prioritizing real-world quality over minimizing mathematical error.
Removing this term, as seen in \cref{tab:ablation}, significantly degrades the quality of the reconstruction audio.

\textbf{Loss Clamp Mechanism:}
During training, we clamp the weights of different loss components to prevent any single term from dominating the gradient updates.
Disabling this mechanism destabilized the training process, causing a substantial drop in performance and even preventing the model from fitting the data effectively.
This demonstrates that loss clamping is essential for balancing competing objectives and ensuring robust convergence.

In summary, these ablation studies confirm that \system's architectural components and training procedures are not just beneficial but essential for achieving SOTA lossless audio reconstruction.